\documentclass[12pt]{article}
\usepackage[latin1]{inputenc}
\usepackage[british]{babel}
\usepackage{cmap}
\usepackage{lmodern}

\usepackage{amssymb, amsmath, amsthm}
\usepackage[a4paper,top=25mm,bottom=25mm,left=25mm,right=25mm]{geometry}
\usepackage{ragged2e}

\usepackage{authblk} 
\usepackage{pifont}
\usepackage{graphicx}
\usepackage[usenames,dvipsnames,svgnames,table]{xcolor}
\usepackage[figuresright]{rotating}
\usepackage{xtab} 
\usepackage{longtable} 
\usepackage{multirow}
\usepackage{footnote}
\usepackage[stable]{footmisc}
\usepackage{chngpage} 
\usepackage{pdflscape} 
\usepackage[nottoc,notlot,notlof]{tocbibind} 

\usepackage{pgfplots}
\pgfplotsset{compat=1.14}
\pgfplotsset{every tick label/.append style={font=\footnotesize}}
\usepgfplotslibrary{fillbetween}
\usepackage{setspace}

\makesavenoteenv{tabular}
\usepackage{tabularx}
\usepackage{booktabs}
\usepackage{threeparttable}
\usepackage[referable]{threeparttablex} 
\newcolumntype{R}{>{\raggedleft\arraybackslash}X}
\newcolumntype{L}{>{\raggedright\arraybackslash}X}
\newcolumntype{C}{>{\centering\arraybackslash}X}
\newcolumntype{M}[1]{>{\centering\arraybackslash}m{#1}}
\newcolumntype{K}{>{\columncolor{gray!20}}C}
\newcolumntype{k}{>{\columncolor{gray!20}}c}

\newlength{\tablen}

\usepackage{dcolumn} 
\newcolumntype{.}{D{.}{.}{-1}}

\usepackage{tikz}
\usetikzlibrary{arrows, calc, matrix, patterns, positioning, trees}
\usepackage[semicolon]{natbib}
\usepackage[hyphens]{url}
\usepackage{hyperref} 
\hypersetup{
  colorlinks   = true,    
  urlcolor     = blue,    
  linkcolor    = blue,    
  citecolor    = ForestGreen      
}
\usepackage{microtype}
\usepackage[justification=centerfirst]{caption} 

\usepackage[labelformat=simple]{subcaption}

\DeclareCaptionLabelFormat{parenthesis}{(#2)}
\makeatletter
\renewcommand\p@subfigure{\arabic{figure}.}
\makeatother

\DeclareCaptionLabelFormat{parenthesis}{(#2)}
\makeatletter
\renewcommand\p@subtable{\arabic{table}.}
\makeatother

\usepackage{enumitem}
\setlist[itemize]{leftmargin=2.5\parindent}
\setlist[enumerate]{leftmargin=2.5\parindent}

\def\addlegendimage{\csname pgfplots@addlegendimage\endcsname}

\theoremstyle{plain}

\theoremstyle{definition}

\newtheorem{definition}{Definition}

\theoremstyle{remark}

\def\keywords{\vspace{.5em} 
{\noindent \textit{Keywords}: }}

\def\JEL{\vspace{.5em} 
{\noindent \textbf{\emph{JEL} classification number}: }}

\def\AMS{\vspace{.5em} 
{\noindent \textbf{\emph{MSC} class}: }}

\author{
\href{https://sites.google.com/view/laszlocsato}{L\'aszl\'o Csat\'o}\thanks{~Corresponding author. E-mail: \emph{laszlo.csato@sztaki.hu}}}
\affil{Institute for Computer Science and Control (SZTAKI) \\
E\"otv\"os Lor\'and Research Network (ELKH) \\
Laboratory on Engineering and Management Intelligence, Research Group of Operations Research and Decision Systems}
\affil{Corvinus University of Budapest (BCE) \\
Department of Operations Research and Actuarial Sciences}
\affil{Budapest, Hungary}

\title{Quantifying the unfairness of the \\ 2018 FIFA World Cup qualification}
\date{\today}

\def\Dedication{ 
{\noindent $\mathfrak{Aber}$ $\mathfrak{so}$ $\mathfrak{viel}$ $\mathfrak{ist}$ $\mathfrak{an}$ $\mathfrak{sich}$ $\mathfrak{klar}$, $\mathfrak{da\ss}$ $\mathfrak{dieser}$, $\mathfrak{wie}$ $\mathfrak{jeder}$ $\mathfrak{Gegenstand}$, $\mathfrak{der}$ $\mathfrak{unser}$ \linebreak $\mathfrak{Begreifungsverm\ddot{o}gen}$ $\mathfrak{nicht}$ $\mathfrak{\ddot{u}bersteigt}$, $\mathfrak{durch}$ $\mathfrak{einen}$ $\mathfrak{untersuchenden}$ $\mathfrak{Geist}$ $\mathfrak{aufgehellt}$ $\mathfrak{und}$ $\mathfrak{in}$ $\mathfrak{seinem}$ $\mathfrak{inneren}$ $\mathfrak{Zusammenhang}$ $\mathfrak{mehr}$ $\mathfrak{oder}$ $\mathfrak{weniger}$ $\mathfrak{deutlich}$ $\mathfrak{gemacht}$ $\mathfrak{werden}$ $\mathfrak{kann}$, $\mathfrak{und}$ $\mathfrak{das}$ $\mathfrak{allein}$ $\mathfrak{reicht}$ $\mathfrak{schon}$ $\mathfrak{hin}$, $\mathfrak{den}$ $\mathfrak{Begriff}$ $\mathfrak{der}$ $\mathfrak{Theorie}$ $\mathfrak{zu}$ $\mathfrak{verwirklichen}$.}\footnote{~``\emph{But so much is evident in itself, that this, like every other subject which does not surpass our powers of understanding, may be lighted up, and be made more or less plain in its inner relations by an enquiring mind, and that alone is sufficient to realise the idea of a \emph{theory}.''}
(Source: Carl von Clausewitz: \emph{On War}, Book 2, Chapter 3---Art or Science of War, translated by Colonel James John Graham, London, N. Tr\"ubner, 1873. \url{http://clausewitz.com/readings/OnWar1873/TOC.htm})}

\flushright
\noindent (Carl von Clausewitz: \emph{Vom Kriege})

\vspace{1cm} 
\justify }

\begin{document}
\newgeometry{top=20mm,bottom=20mm,left=25mm,right=25mm}

\maketitle
\thispagestyle{empty}
\Dedication

\begin{abstract}
\noindent
This paper investigates the fairness of the 2018 FIFA World Cup qualifying competition via Monte Carlo simulations. The qualifying probabilities are calculated for 102 nations, all teams except for African and European countries. A method is proposed to quantify the degree of unfairness. Although the qualifications within four FIFA confederations are constructed fairly, serious differences are found between the continents: for instance, a South American team could have tripled its chances by playing in Asia. Choosing a fixed matchup in the inter-continental play-offs instead of the current random draw can reduce the unfairness of the competition. The move of Australia from the Oceanian to the Asian zone is shown to increase its probability of participating in the 2018 FIFA World Cup by about 65\%. Our results provide important insights for the administrators on how to reallocate the qualifying berths.


\AMS{62F07, 68U20}

\JEL{C44, C63, Z20}
\end{abstract}

\clearpage
\newgeometry{top=25mm,bottom=25mm,left=25mm,right=25mm}

\section{Introduction} \label{Sec1}

The FIFA World Cup, the most prestigious soccer tournament around the world, is followed by millions of fans. According to \citet{Palacios-Huerta2014}, 5\% of all the people who \emph{ever} lived on the Earth watched the final of the 2010 FIFA World Cup played by the Netherlands and Spain. Qualification to the FIFA World Cup creates widespread media coverage in the competing countries \citep{FrawleyVandenHoven2015} and brings significant economic benefits \citep{StoneRod2016}: each participating team has received at least 9.5 million USD in the 2018 FIFA World Cup \citep{FIFA2017b}.
Success in soccer can even help build nations \citep{Depetris-ChauvinDuranteCampante2020}.

Obviously, every national team cannot play in the FIFA World Cup. Thus there is a World Cup qualification competition, consisting of a series of tournaments organised by the six FIFA confederations.
In 2006, Australia left the Oceania Football Confederation (OFC) to join the Asian Football Confederation (AFC). Since then, the country qualified for all FIFA World Cups (2010, 2014, 2018), although it played only twice before (1974, 2006). Is this a mere coincidence, the result of a fruitful team development strategy, or can be partially explained by the change of affiliation?
The current paper discusses the third issue by analysing the fairness of the 2018 FIFA World Cup qualification.

Some scientific research has addressed the World Cup qualifiers.
\citet{Flegl2014} applies Data Envelopment Analysis (DEA) to evaluate the performance of national soccer teams during the 2014 FIFA World Cup qualification.
\citet{StoneRod2016} aim to assess the degree to which the allocation of qualification berths among the six FIFA confederations reflects the quality of the teams from their specific region, mainly by descriptive statistics.
\citet{DuranGuajardoSaure2017} recommend integer programming to construct alternative schedules for the South American qualifiers that overcome the main drawbacks of the previous approach. Their proposal has been unanimously approved by the South American associations to use in the 2018 FIFA World Cup qualification.
\citet{PollardArmatas2017} investigate home advantage in the FIFA World Cup qualification games. 
\citet{Csato2020c} identifies an incentive incompatibility problem in the European section of the recent FIFA World Cup qualifications.
 
However, the FIFA World Cup qualification competition has never been studied before with respect to the chances of the participating teams. A possible reason is the complexity of the qualifying system as will be seen later.

Our paper aims to fill this research gap.
In particular, the probability of qualification to the 2018 FIFA World Cup is quantified for the 102 nations of the AFC, CONCACAF (Confederation of North, Central American and Caribbean Association Football), CONMEBOL (South American Football Confederation), and OFC to answer three questions:
(a) Is the qualification process designed \emph{fairly} in the sense that it provides a higher chance for a better team both within and between the confederations?
(b) Is it possible to improve fairness without reallocating the qualifying berths?
(c) How did the move of Australia from the OFC to the AFC in 2006 affect the teams?

The main contributions can be summarised as follows:
\begin{enumerate}
\item
First in the academic literature, the paper calculates the qualifying probabilities for the FIFA World Cup based on the Elo ratings of the national teams.
\item
A method is proposed to measure the degree of unfairness. It shows that all the four qualifiers are constructed fairly. On the contrary, substantial differences are found between the confederations.
\item
Using a well-devised fixed matchup in the inter-continental play-offs---a policy applied in the 2010 FIFA World Cup qualification---instead of the current random draw can reduce unfairness by 7.5\%.
\item
Australia has increased its probability of playing in the 2018 FIFA World Cup by 65\% as a result of leaving the OFC and joining the AFC. The move has been detrimental to any AFC nation, while it has favoured all other countries, especially New Zealand. The change has reduced unfairness by almost 20\%, which might explain why the move of Australia was approved.
\end{enumerate}

Our approach can be applied in any sports where teams contest in parallel tournaments for the same prize, hence the natural issue of equal treatment of equals emerges. While these designs have recently been analysed with respect to incentive incompatibility \citep{Vong2017, DagaevSonin2018, Csato2020d}, the numerical investigation of fairness is currently limited to the qualification for the UEFA Euro 2020 \citep[Chapter~6.4]{Csato2021a}.


\section{Related literature} \label{Sec2}

The paper contributes to three different fields of research: fairness in sports, analysis of FIFA competitions and rules, and simulation of tournament designs.

Fairness in sports is a widely discussed topic in the literature. From the perspective of tournament design, there are two crucial axioms:
(1) stronger players should be preferred to weaker players; and
(2) equal players should be treated equally.
Otherwise, an incentive might exist to manipulate the tournament.
\citet{GrohMoldovanuSelaSunde2012} check which seedings in elimination tournaments satisfy the two properties.
\citet{ArlegiDimitrov2020} apply these requirements to different kinds of knockout contests and characterise the appropriate structures.
\citet[Chapter~6.4]{Csato2021a} shows the unfairness of the qualification for the 2020 UEFA European Championship with important lessons for sports management \citep{HaugenKrumer2021}.
Both theoretical \citep{KrumerMegidishSela2017a, KrumerMegidishSela2020a, LaicaLauberSahm2021, Sahm2019} and empirical \citep{KrumerLechner2017} investigations reveal that the ex-ante winning probabilities in round-robin tournaments with three and four symmetric players may depend on the schedule, which can lead to severe problems in the 2026 FIFA World Cup \citep{Guyon2020a}.
Soccer penalty shootouts seem to be advantageous for the first shooter \citep{ApesteguiaPalacios-Huerta2010, Palacios-Huerta2014, VandebroekMcCannVroom2018} but this bias can be mitigated by a carefully devised mechanism \citep{AnbarciSunUnver2021, BramsIsmail2018, Csato2021c, CsatoPetroczy2022, LambersSpieksma2021, Palacios-Huerta2012}.
The knockout bracket of the 2016 UEFA European Championship has created imbalances among the six round-robin groups of four teams each \citep{Guyon2018a}.
\citet{VaziriDabadghaoYihMorin2018} state some fairness properties of sports ranking methods.

In contrast to the World Cup qualifiers, several attempts have been made to forecast the FIFA World Cup final tournament games.
\citet{DyteClarke2000} treat the goals scored by the teams as independent Poisson variables to simulate the 1998 FIFA World Cup.
\citet{Deutsch2011} aims to judge the impact of the draw in the 2010 World Cup, as well as to look back and identify surprises, disappointments, and upsets.
\citet{GrollSchaubergerTutz2015} fit and examine two models to forecast the 2014 FIFA World Cup.
\citet{OLeary2017} finds that the Yahoo crowd was statistically significantly better at predicting the outcomes of matches in the 2014 World Cup compared to the experts and was similar in performance to established betting odds.
\citet{GrollLeySchaubergerVanEetvelde2019} propose a novel hybrid modelling approach by combining random forests with Poisson ranking methods to predict the 2018 FIFA World Cup. This technique has also been used to forecast the 2019 FIFA Women's World Cup \citep{GrollLeySchaubergerVanEetveldeZeileis2019} and the 2020 UEFA European Championship \citep{GrollHvattumLeyPoppSchaubergerVanEetveldeZeileis2021}.

Further aspects of the FIFA World Cup have also been researched extensively.
\citet{Jones1990} and \citet{RathgeberRathgeber2007} discuss the consequences of the unevenly distributed draw procedures for the 1990 and 2006 FIFA World Cups, respectively.
\citet{ScarfYusof2011} reveal the effect of seeding policy and other design changes on the progression of competitors in the World Cup final tournament.
\citet{Guyon2015a} collects some flaws and criticisms of the World Cup seeding system. \citet{LalienaLopez2019} and \citet{CeaDuranGuajardoSureSiebertZamorano2020} provide a detailed analysis of group assignment in the FIFA World Cup.

Finally, since historical data usually do not make it possible to calculate the majority of tournament metrics such as qualifying probabilities, it is necessary to use computer simulations for this purpose, especially for evaluating new designs \citep{ScarfYusofBilbao2009}.
Any simulation model should be based on a prediction model for individual ties.
According to \citet{LasekSzlavikBhulai2013}, the best performing algorithm of ranking systems in soccer with respect to accuracy is a version of the famous Elo rating.
\citet{BakerMcHale2018} provide time-varying ratings for international soccer teams.
\citet{VanEetveldeLey2019} overview the most common ranking methods in soccer.
\citet{LeyVandeWieleVanEeetvelde2019} build a ranking reflecting the teams' current strengths and illustrate its usefulness by examples where the existing rankings fail to provide enough information or lead to peculiar results.
\citet{CoronaForrestTenaWiper2019} propose a Bayesian approach to take into
account the uncertainty of parameter estimates in the underlying match-level
forecasting model.

\section{The 2018 FIFA World Cup qualification} \label{Sec3}

The \href{https://en.wikipedia.org/wiki/FIFA_World_Cup_qualification}{FIFA World Cup qualification} is a series of tournaments to determine the participants of the \href{https://en.wikipedia.org/wiki/FIFA_World_Cup}{FIFA World Cup}. Since 1998, the final competition contains 32 teams such that the host nation(s) receive(s) a guaranteed slot.
The number of qualifying berths for the continents is fixed from 2006 to 2022 as follows:
\begin{itemize}
\item
AFC (Asian Football Confederation): 4.5;
\item
CAF (Confederation of African Football): 5;
\item
CONCACAF (Confederation of North, Central American and Caribbean Association Football): 3.5;
\item
CONMEBOL (South American Football Confederation): 4.5;
\item
OFC (Oceania Football Confederation): 0.5;
\item
UEFA (Union of European Football Associations): 13.
\end{itemize}
The six confederations organise their own contests.
The 0.5 slots represent a place in the inter-continental play-offs, which is the only interaction between the qualifying tournaments of different geographical zones.

The qualifications of all confederations are played in rounds. Each round is designed either in a \emph{knockout} format (where two teams play two-legged home-away matches) or in a \emph{round-robin} format (where more than two teams play in a single or home-away group against every other team of the group). The rounds are often \emph{seeded}, that is, the participating countries are divided into the same number of pots as the number of teams per group (meaning two pots in the knockout format) and one team from each pot goes to a given group. The \emph{traditional seeding} is based on an exogenously given ranking---usually the FIFA World Ranking at a specific date---such that, if a pot contains $k$ teams, the best $k$ teams are in the first pot, the next $k$ are in the second pot, and so on.

Our paper focuses on four qualifications, the AFC, the CONCACAF, the CONMEBOL, and the OFC because
(1) contrary to the CAF and UEFA competitions, they are connected to each other;
(2) the largest and most successful nation of the OFC, Australia, switched to the AFC in 2006.

The \href{https://en.wikipedia.org/wiki/2018_FIFA_World_Cup_qualification_(AFC)}{2018 FIFA World Cup qualification (AFC)} contained $46$ nations and four rounds.
The starting access list was determined by the FIFA World Ranking of January 2015.
\begin{itemize}
\item
\href{https://en.wikipedia.org/wiki/2018_FIFA_World_Cup_qualification_\%E2\%80\%93_AFC_First_Round}{\textbf{First round}} \\
Format: knockout \\
Competitors: the $12$ lowest-ranked teams ($35$--$46$) \\
Seeding: traditional; based on the FIFA World Ranking of January 2015
\item
\href{https://en.wikipedia.org/wiki/2018_FIFA_World_Cup_qualification_\%E2\%80\%93_AFC_Second_Round}{\textbf{Second round}} \\
Format: home-away round-robin, 8 groups of five teams each \\
Competitors: the $34$ highest-ranked teams ($1$--$34$) + the six winners from the first round \\
Seeding: traditional; based on the FIFA World Ranking of April 2015\footnote{~Since the seeding order differed from the ranking in the AFC entrant list, three winners in the first round (India, Timor-Leste, Bhutan) were not seeded in the weakest pot 5.}
\item
\href{https://en.wikipedia.org/wiki/2018_FIFA_World_Cup_qualification_\%E2\%80\%93_AFC_Third_Round}{\textbf{Third round}} \\
Format: home-away round-robin, 2 groups of six teams each \\
Competitors: the eight group winners and the four best runners-up in the second round\footnote{~Group F in the second round consisted of only four teams because Indonesia was disqualified by FIFA. Therefore, the matches played against the fifth-placed team were disregarded in the comparison of the runners-up.} \\
Seeding: traditional; based on the FIFA World Ranking of April 2016 \\
The two group winners and the two runners-up qualified to the 2018 FIFA World Cup.
\item
\href{https://en.wikipedia.org/wiki/2018_FIFA_World_Cup_qualification_\%E2\%80\%93_AFC_Fourth_Round}{\textbf{Fourth round}} \\
Format: knockout \\
Competitors: the third-placed teams from the groups in the third round \\
Seeding: redundant \\
The winner advanced to the \href{https://en.wikipedia.org/wiki/2018_FIFA_World_Cup_qualification_(inter-confederation_play-offs)}{inter-confederation play-offs}.
\end{itemize}

The \href{https://en.wikipedia.org/wiki/2018_FIFA_World_Cup_qualification_(CONCACAF)}{2018 FIFA World Cup qualification (CONCACAF)} contained $35$ nations and five rounds.
The access list was determined by the FIFA World Ranking of August 2014.
\begin{itemize}
\item
\href{https://en.wikipedia.org/wiki/2018_FIFA_World_Cup_qualification_\%E2\%80\%93_CONCACAF_First_Round}{\textbf{First round}} \\
Format: knockout \\
Competitors: the $14$ lowest-ranked teams ($22$--$35$) \\
Seeding: traditional; based on the FIFA World Ranking of August 2014
\item
\href{https://en.wikipedia.org/wiki/2018_FIFA_World_Cup_qualification_\%E2\%80\%93_CONCACAF_Second_Round}{\textbf{Second round}} \\
Format: knockout \\
Competitors: the teams ranked $9$--$21$ in the access list + the seven winners from the first round \\
Seeding: the seven teams of pot 5 (ranked $9$--$15$) were drawn against the teams of pot 6 (the winners from the first round) and the three teams of pot 3 (ranked $16$--$18$) were drawn against the three teams of pot 4 (ranked $19$--$21$); based on the FIFA World Ranking of August 2014
\item
\href{https://en.wikipedia.org/wiki/2018_FIFA_World_Cup_qualification_\%E2\%80\%93_CONCACAF_Third_Round}{\textbf{Third round}} \\
Format: knockout \\
Competitors: the teams ranked $7$--$8$ in the access list + the $10$ winners from the second round \\
Seeding: traditional; based on the FIFA World Ranking of August 2014
\item
\href{https://en.wikipedia.org/wiki/2018_FIFA_World_Cup_qualification_\%E2\%80\%93_CONCACAF_Fourth_Round}{\textbf{Fourth round}} \\
Format: home-away round-robin, 3 groups of four teams each \\
Competitors: the teams ranked $1$--$6$ in the access list + the six winners from the third round \\
Seeding: pot 1 (teams ranked $1$--$3$), pot 2 (teams ranked $4$--$6$), pot 3 (the winners from the third round) such that each group contained a team from pot 1, a team from pot 2, and two teams from pot 3; based on the FIFA World Ranking of August 2014
\item
\href{https://en.wikipedia.org/wiki/2018_FIFA_World_Cup_qualification_\%E2\%80\%93_CONCACAF_Fifth_Round}{\textbf{Fifth round}} \\
Format: home-away round-robin, one group of six teams \\
Competitors: the group winners and the runners-up in the fourth round \\
Seeding: redundant \\
The top three teams qualified to the 2018 FIFA World Cup and the fourth-placed team advanced to the \href{https://en.wikipedia.org/wiki/2018_FIFA_World_Cup_qualification_(inter-confederation_play-offs)}{inter-confederation play-offs}.
\end{itemize}
The CONCACAF competition can be criticised for the great role attributed to the FIFA World Ranking. Perhaps it is not only a coincidence that this confederation has planned to fundamentally restructure its \href{https://en.wikipedia.org/wiki/2022_FIFA_World_Cup_qualification_(CONCACAF)}{qualifying tournament for the 2022 FIFA World Cup} \citep{CONCACAF2019}.

The \href{https://en.wikipedia.org/wiki/2018_FIFA_World_Cup_qualification_(CONMEBOL)}{2018 FIFA World Cup qualification (CONMEBOL)} contained $10$ nations, which contested in a home-away round-robin tournament \citep{DuranGuajardoSaure2017}. The top four teams qualified to the 2018 FIFA World Cup, and the fifth-placed team advanced to the \href{https://en.wikipedia.org/wiki/2018_FIFA_World_Cup_qualification_(inter-confederation_play-offs)}{inter-confederation play-offs}.

The \href{https://en.wikipedia.org/wiki/2018_FIFA_World_Cup_qualification_(OFC)}{2018 FIFA World Cup qualification (OFC)} contained $11$ nations and four rounds.
\begin{itemize}
\item
\href{https://en.wikipedia.org/wiki/2018_FIFA_World_Cup_qualification_\%E2\%80\%93_OFC_First_Round}{\textbf{First round}} \\
Format: single round-robin, one group organised in a country (Tonga was chosen later) \\
Competitors: the four lowest-ranked teams ($8$--$11$), based on FIFA World Ranking and sporting reasons \\
Seeding: redundant
\item
\href{https://en.wikipedia.org/wiki/2018_FIFA_World_Cup_qualification_\%E2\%80\%93_OFC_Second_Round}{\textbf{Second round}} \\
Format: single round-robin, 2 groups of four teams each, all matches played in one country \\
Competitors: the seven strongest teams ($1$--$7$) + the group winner in the first round \\
Seeding: traditional; based on the FIFA World Ranking of July 2015
\item
\href{https://en.wikipedia.org/wiki/2018_FIFA_World_Cup_qualification_\%E2\%80\%93_OFC_Third_Round}{\textbf{Third round}} \\
Format: home-away round-robin, 2 groups of three teams each \\
Competitors: the top three teams from each group in the second round \\
Seeding: pot 1 (\href{https://en.wikipedia.org/wiki/2016_OFC_Nations_Cup}{2016 OFC Nations Cup} finalists), pot 2 (\href{https://en.wikipedia.org/wiki/2016_OFC_Nations_Cup}{2016 OFC Nations Cup} semifinalists), pot 3 (third-placed teams in the second round) such that each group contained one team from pots $1$--$3$ each\footnote{~The group stage of the \href{https://en.wikipedia.org/wiki/2016_OFC_Nations_Cup}{2016 OFC Nations Cup} served as the second round of the 2018 FIFA World Cup qualification (OFC). Any group winner was matched with the runner-up of the other group in the semifinals of the 2016 OFC Nations Cup.}
\item
\href{https://en.wikipedia.org/wiki/2018_FIFA_World_Cup_qualification_\%E2\%80\%93_OFC_Third_Round#Final}{\textbf{Fourth round}} \\
Format: knockout \\
Competitors: the group winners in the third round \\
Seeding: redundant \\
The winner advanced to the \href{https://en.wikipedia.org/wiki/2018_FIFA_World_Cup_qualification_(inter-confederation_play-offs)}{inter-confederation play-offs}.
\end{itemize}

Consequently, the \href{https://en.wikipedia.org/wiki/2018_FIFA_World_Cup_qualification_(inter-confederation_play-offs)}{inter-confederation play-offs} were contested by four teams from the four confederations (AFC, CONCACAF, CONMEBOL, OFC), and were played in a knockout format. The four nations were drawn randomly into two pairs without seeding. The two winners qualified to the 2018 FIFA World Cup.
The inter-confederation play-offs of the \href{https://en.wikipedia.org/wiki/2006_FIFA_World_Cup_qualification_(inter-confederation_play-offs)}{2006 FIFA World Cup} and the \href{https://en.wikipedia.org/wiki/2014_FIFA_World_Cup_qualification_(inter-confederation_play-offs)}{2014 FIFA World Cup qualification} were also drawn randomly. This policy will be followed in the \href{https://en.wikipedia.org/wiki/2022_FIFA_World_Cup_qualification_(inter-confederation_play-offs)}{2022 FIFA World Cup}, too.
However, FIFA fixed the ties in the \href{https://en.wikipedia.org/wiki/2010_FIFA_World_Cup_qualification_(inter-confederation_play-offs)}{inter-continental play-offs of the 2010 FIFA World Cup qualification} as AFC vs.\ OFC and CONCACAF vs.\ CONMEBOL to pair teams being in closer time zones.\footnote{~Similarly, FIFA matched a randomly drawn UEFA runner-up with the AFC team, and two nations from CONMEBOL and OFC in the two \href{https://en.wikipedia.org/wiki/2002_FIFA_World_Cup_qualification_(inter-confederation_play-offs)}{inter-continental play-offs of the 2002 FIFA World Cup qualification}.}

\section{Methodology and implementation} \label{Sec4}

This section presents how the outcome of a qualifying match is determined, the simulation is carried out, and the unfairness of a qualification tournament is measured.

\subsection{The simulation model} \label{Sec41}

In order to quantify a particular tournament metric of the FIFA World Cup qualification, it is necessary to follow a simulation technique because historical data are limited: the national teams do not play many matches and the outcome of a qualification is only a single realisation of several random variables. Such a model should be based on predicting the result of individual games. For this purpose, the strengths of the teams are measured by the \href{https://en.wikipedia.org/wiki/World_Football_Elo_Ratings}{World Football Elo Ratings}, available at the website \href{http://eloratings.net/}{eloratings.net}.
Elo-inspired methods are usually good in forecasting \citep{LasekSzlavikBhulai2013}, and have been widely used in academic research \citep{HvattumArntzen2010, LasekSzlavikGagolewskiBhulai2016, CeaDuranGuajardoSureSiebertZamorano2020}.

Elo ratings depend on the results of previous matches but the same result is worth more when the opponent is stronger. Furthermore, playing new games decreases the weight of earlier matches. Since there is no official Elo rating for national teams, this approach can be implemented in various ways. For instance, while the official \href{https://en.wikipedia.org/wiki/FIFA_World_Rankings}{FIFA World Ranking} adopted the Elo method of calculation after the 2018 FIFA World Cup \citep{FIFA2018a, FIFA2018c}, it does not contain any adjustment for home or away games. However, home advantage has been presented to be a crucial factor in international soccer, even though its influence appears to be narrowing \citep{BakerMcHale2018}.
The World Football Elo Ratings takes into account some soccer-specific parameters such as the margin of victory, home advantage, and the tournament where the match was played.

In the 2018 FIFA World Cup qualification, three types of matches were played: group matches in a home-away format, single group matches in a randomly chosen country that is assumed to be a neutral field (only in the first and the second rounds in the OFC zone), and home-away knockout matches.
The win expectancy of team $i$ against team $j$ can be directly obtained from the formula of Elo rating according to the system of World Football Elo Ratings (see \href{http://eloratings.net/about}{http://eloratings.net/about}):
\begin{equation} \label{eq1}
W_{ij} = \frac{1}{1 + 10^{-(E_i - E_j)/400}},
\end{equation}
where $E_i$ and $E_j$ are the Elo ratings of the two teams, respectively, such that the rating of the home team is increased by $100$.

In knockout clashes, the teams focus primarily on advancing to the next round rather than winning one match. Therefore, we have followed the solution of the ClubElo rating (see \href{http://clubelo.com/System}{http://clubelo.com/System}), namely, these two-legged matches are considered as one long match with a corresponding increase in the difference between the strengths of the teams. The probability that team $i$ advances against team $j$ is:
\begin{equation} \label{eq2}
W_{ij}^{(KO)} = \frac{1}{1 + 10^{- \sqrt{2} (E_i - E_j)/400}}.
\end{equation}

In the case of group matches, the number of goals scored by each team in each match should also be modelled because drawn matches are common in football, and ties between teams with the same score are decided by goal difference. For this purpose, the number of goals is assumed to follow a Poisson distribution, which is a standard solution in the literature, see, e.g.\ \citet{ChaterArrondelGayantLaslier2021} or \citet{DagaevRudyak2019}. In particular, team $i$ scores $k$ goals against team $j$ with the probability
\begin{equation} \label{Poisson_dist}
P_{ij}(k) = \frac{ \left( \lambda_{ij}^{(f)} \right)^k \exp \left( -\lambda_{ij}^{(f)} \right)}{k!},
\end{equation}
where the parameter $\lambda_{ij}^{(f)}$ is the expected number of goals scored by team $i$ against team $j$ if the match is played on field $f$ (home: $f = h$; away: $f = a$; neutral $f = n$).

The value of parameter $\lambda_{ij}^{(f)}$ in \eqref{Poisson_dist} is modelled as a quartic polynomial of $W_{ij}$ (see \eqref{eq1}) estimated using a least squares regression with a regime change at $W_{ij} = 0.9$ \citep{FootballRankings2020}.
On the basis of more than 29 thousand home-away matches played by national football teams, the average expected number of goals for the home team $i$ is
\begin{equation} \label{Exp_goals_home}
\lambda_{ij}^{(h)} = 
\left\{ \begin{array}{ll}
-5.42301 \cdot W_{ij}^4 + 15.49728 \cdot W_{ij}^3 \\
- 12.6499 \cdot W_{ij}^2 + 5.36198 \cdot W_{ij} + 0.22862 & \textrm{if } W_{ij} \leq 0.9 \\ \\
231098.16153 \cdot (W_{ij}-0.9)^4 - 30953.10199 \cdot (W_{ij}-0.9)^3 & \\
+ 1347.51495 \cdot (W_{ij}-0.9)^2 - 1.63074 \cdot (W_{ij}-0.9) + 2.54747 & \textrm{if } W_{ij} > 0.9
\end{array} \right.
\end{equation}
with $R^2 = 0.984$, whereas the average number of goals for the away team $j$ equals
\begin{equation} \label{Exp_goals_away}
\lambda_{ij}^{(a)} = 
\left\{ \begin{array}{ll}
90173.57949 \cdot (W_{ij} - 0.1)^4 + 10064.38612 \cdot (W_{ij} - 0.1)^3 \\
+ 218.6628 \cdot (W_{ij} - 0.1)^2 - 11.06198 \cdot (W_{ij} - 0.1) + 2.28291 & \textrm{if } W_{ij} < 0.1 \\ \\
-1.25010 \cdot W_{ij}^4 -  1.99984 \cdot W_{ij}^3 & \\
+ 6.54946 \cdot W_{ij}^2 - 5.83979 \cdot W_{ij} + 2.80352 & \textrm{if } W_{ij} \geq 0.1
\end{array} \right.
\end{equation}
with $R^2 = 0.955$.

In the OFC zone, some matches are played on neutral ground when, based on almost 10 thousand matches, the average number of goals for team $i$ is
\begin{equation} \label{Exp_goals_neutral}
\lambda_{ij}^{(n)} = 
\left\{ \begin{array}{ll}
3.90388 \cdot W_{ij}^4 - 0.58486 \cdot W_{ij}^3 \\
- 2.98315 \cdot W_{ij}^2 + 3.13160 \cdot W_{ij} + 0.33193 & \textrm{if } W_{ij} \leq 0.9 \\ \\
308097.45501 \cdot (W_{ij}-0.9)^4 - 42803.04696 \cdot (W_{ij}-0.9)^3 & \\
+ 2116.35304 \cdot (W_{ij}-0.9)^2 - 9.61869 \cdot (W_{ij}-0.9) + 2.86899 & \textrm{if } W_{ij} > 0.9,
\end{array} \right.
\end{equation}
with $R^2 = 0.976$.

The ranking of the teams in a group is determined as follows:
(a) greatest number of points obtained in all group matches;
(b) goal difference in all group matches;
(c) greatest number of goals scored in all group matches;
(d) drawing of lots.

The Elo ratings are dynamic but we have fixed them for the sake of simplicity. In each of the four confederations, the ratings of all teams on the day before the first match of the relevant qualification tournament are considered. Four tables in the Appendix show the corresponding measures of strength:
Table~\ref{Table_A1} for the $35$ CONCACAF teams;
Table~\ref{Table_A2} for the $46$ AFC teams;
Table~\ref{Table_A3} for the $10$ CONMEBOL teams; and
Table~\ref{Table_A4} for the $11$ OFC teams.

Any theoretical model is only as good as its assumptions. Therefore, it is worth summarising the main limitations:
\begin{itemize}
\item
The strength of the teams is exogenously given and fixed during the whole qualification process.
\item
Home advantage does not differ between the confederations despite the findings of \citet{PollardArmatas2017}. However, the influence of the corresponding parameter is minimal since all matches are played both home and away except for Oceania, where some games are hosted by a randomly drawn country.
\item
The schedule of the matches is not accounted for.
\item
The efforts of the teams do not change even if they have already qualified or have been eliminated. 
\end{itemize}
These shortcomings mean that the numerical results are primarily for comparative purposes.

\subsection{Implementation} \label{Sec42}

Our computer code closely follows the rules of the qualification process described in the previous section. Pot assignment and seeding are based on the Elo rating of the teams in each case since the strengths of the teams are given by this measure instead of the FIFA ranking, which was a rather bad predictor before the 2018 reform \citep{LasekSzlavikBhulai2013, CeaDuranGuajardoSureSiebertZamorano2020}.
Although the official AFC qualification updated the ranking of the teams before the seeding in each round---thus the results of matches played already during the qualification may have affected the subsequent rounds---, that complication is disregarded in our simulations.
Using the Elo rating for seeding is also necessary to guarantee the consistency of the simulations since the proposed fairness measure will be based on the Elo ratings. Therefore, every instance of unfairness will be intrinsic to the tournament design.

Finally, the move of Australia from the OFC to the AFC will also be evaluated. Accordingly, an alternative design of the FIFA World Cup qualification should be chosen with Australia being in the OFC instead of the AFC. Since then there are only $45$ countries in Asia, a straightforward format would be to organise the first knockout round with the 10 lowest-ranked teams ($36$--$45$), while the second round is contested by the $35$ highest-ranked teams ($1$--$35$) plus the five winners from the first round.
Together with Australia, the OFC qualification contains $12$ teams. Fortunately, the design of the \href{https://en.wikipedia.org/wiki/2006_FIFA_World_Cup_qualification_(OFC)}{2006 FIFA World Cup qualification (OFC)} can be adopted without any changes:
\begin{itemize}
\item
\href{https://en.wikipedia.org/wiki/2006_FIFA_World_Cup_qualification_\%E2\%80\%93_OFC_First_Round}{\textbf{First round}} \\
Format: single round-robin, 2 groups of five teams each, held in one country \\
Competitors: the $10$ lowest-ranked teams, that is, all nations except for Australia and New Zealand \\
Seeding: traditional\footnote{~This is only a (reasonable) conjecture as we have not found the official regulation.}
\item
\href{https://en.wikipedia.org/wiki/2004_OFC_Nations_Cup}{\textbf{Second round}} \\
Format: single round-robin, one group of six teams, held in one country \\
Competitors: the two highest-ranked teams (Australia, New Zealand) + the group winners and the runners-up in the first round \\
Seeding: redundant
\item
\href{https://en.wikipedia.org/wiki/2006_FIFA_World_Cup_qualification_(OFC)\#Final_round}{\textbf{Third round}} \\
Format: knockout \\
Competitors: the group winner and the runner-up in the second round \\
Seeding: redundant \\
The winner advanced to the \href{https://en.wikipedia.org/wiki/2006_FIFA_World_Cup_qualification_(inter-confederation_play-offs)}{inter-confederation play-offs}.
\end{itemize}

The simulation has been carried out with 2.5 million independent runs in each scenario (e.g.\ when Australia plays in the OFC zone). The probability of qualification is provided by the mean of these iterations, for example, it equals 60\% if a national team qualifies in 1.5 million runs. A further increase does not reduce statistical errors considerably and would be a futile exercise anyway in view of the warnings above.

\subsection{A reasonable measure of unfairness} \label{Sec43}

The degree of unfairness can be quantified by ranking the teams according to their Elo rating and summing the differences of qualifying probabilities that do not fall into line with this ranking. However, in order to ensure the comparability of different tournaments, it is worth guaranteeing independence of the number of competitors $n$ and the number of qualifying slots $q$.\footnote{~We are grateful to \emph{Dries Goossens} for this remark.}
Since the worst case with respect to fairness is when the $q$ lowest-ranked teams qualify with probability one, the crude sum is divided by its theoretical maximum of $(n-q)q$ to get a normalised measure between zero and one.

\begin{definition} \label{Def1}
\emph{Measure of unfairness}:
\begin{equation} \label{eq_unfairness}
\textit{UF} = \sum_{E_i \geq E_j} \frac{\max \{ 0; p(j) - p(i) \}}{(n-q)q},
\end{equation}
where $E_i$ and $p(i)$ are the Elo rating and the probability of qualification for team $i$, respectively, $n$ is the number of competitors and $q$ is the number of qualifying slots.
\end{definition}
Formula~\eqref{eq_unfairness} only considers the ordinal strength of the teams because prescribing how the differences in Elo rating should be converted into differences in qualifying probabilities would require further assumptions, which seem to be challenging to justify.

\section{Results} \label{Sec5}

The three main research questions, presented in the Introduction, will be discussed separately.

\subsection{Quantifying unfairness}

\begin{figure}[t]
\centering

\begin{tikzpicture}
\begin{axis}[width = 1\textwidth, 
height = 0.6\textwidth,
xmin = 350,
xmax = 2100,
ymin = -0.02,
ymax = 1.02,
ymajorgrids,
xlabel = Elo rating,
xlabel style = {font = \small},
ylabel = The probability of qualification,
xlabel style = {font = \small},
legend entries = {AFC$\qquad$,CONCACAF$\qquad$,CONMEBOL$\qquad$,OFC},
legend style = {at = {(0.5,-0.15)},anchor = north,legend columns = 4,font = \small}
]
\addplot[black,thick,only marks,mark=otimes*, mark size=2pt] coordinates {
(1757,0.794028)
(1736,0.743646)
(1727,0.7083252)
(1724,0.6995076)
(1666,0.4492748)
(1625,0.3146624)
(1606,0.2308672)
(1525,0.0741424)
(1522,0.0546336)
(1509,0.0424052)
(1503,0.0373676)
(1487,0.0270876)
(1478,0.0222912)
(1474,0.0202376)
(1446,0.010896)
(1446,0.0107868)
(1403,0.00194)
(1363,0.0005332)
(1274,0.0000236)
(1262,0.0000104)
(1261,0.0000172)
(1236,0.0000028)
(1225,0.0000016)
(1225,0.0000016)
(1211,0)
(1207,0.0000004)
(1173,0)
(1152,0)
(1100,0)
(1084,0)
(1068,0)
(1068,0)
(1062,0)
(952,0)
(943,0)
(883,0)
(837,0)
(829,0)
(806,0)
(795,0)
(734,0)
(717,0)
(639,0)
(636,0)
(606,0)
(521,0)
};

\addplot[ForestGreen,very thick,only marks,mark=x, mark size=3pt] coordinates {
(1893,0.9681452)
(1866,0.9484876)
(1816,0.8820108)
(1706,0.5144224)
(1539,0.066664)
(1539,0.0670488)
(1525,0.0395648)
(1522,0.0374464)
(1507,0.0270288)
(1507,0.027104)
(1446,0.0075236)
(1396,0.0022232)
(1271,0.0000296)
(1244,0.0000084)
(1235,0.0000048)
(1221,0.0000028)
(1210,0.0000016)
(1204,0.0000016)
(1162,0)
(1158,0.0000004)
(1155,0)
(1138,0)
(1132,0)
(1105,0)
(1087,0)
(1059,0)
(951,0)
(941,0)
(937,0)
(933,0)
(776,0)
(679,0)
(619,0)
(612,0)
(575,0)
};

\addplot[red,thick,only marks,mark=diamond*, mark size=3pt] coordinates {
(2067,0.9879564)
(2048,0.9807432)
(1999,0.9394088)
(1977,0.90422)
(1873,0.5398968)
(1822,0.2924588)
(1807,0.2354568)
(1696,0.0279288)
(1668,0.0137624)
(1636,0.0057576)
};

\addplot[blue,thick,only marks,mark=star, mark size=3pt] coordinates {
(1516,0.1844932)
(1357,0.0340024)
(1281,0.0126864)
(1226,0.0058176)
(1175,0.002478)
(1174,0.0023724)
(1053,0.0001516)
(684,0)
(681,0)
(675,0)
(381,0)
};

\node[pin={[ultra thick, pin distance=0.5cm, pin edge = black, above left, font=\footnotesize]{Japan}}] at (1757,0.794028) {};
\node[pin={[ultra thick, pin distance=0.5cm, pin edge = black, above left, font=\footnotesize]{Australia}}] at (1727,0.7083252) {};


\node[pin={[ultra thick, pin distance=0.5cm, pin edge = ForestGreen, font=\footnotesize]275:{\textcolor{ForestGreen}{\footnotesize{Mexico}}}}] at (1893,0.9681452) {};

\node[pin={[ultra thick, pin distance=0.5cm, pin edge = red, font=\footnotesize]360:{\textcolor{red}{\footnotesize{Peru}}}}] at (1807,0.2354568) {};

\node[pin={[ultra thick, pin distance=0.5cm, pin edge = blue, above left, font=\footnotesize]{\textcolor{blue}{\footnotesize{New Zealand}}}}] at (1516,0.1844932) {};
\end{axis}
\end{tikzpicture}

\caption[The probability of qualification for the 2018 FIFA World Cup]{The probability of qualification for the 2018 FIFA World Cup}
\label{Fig1}

\end{figure}
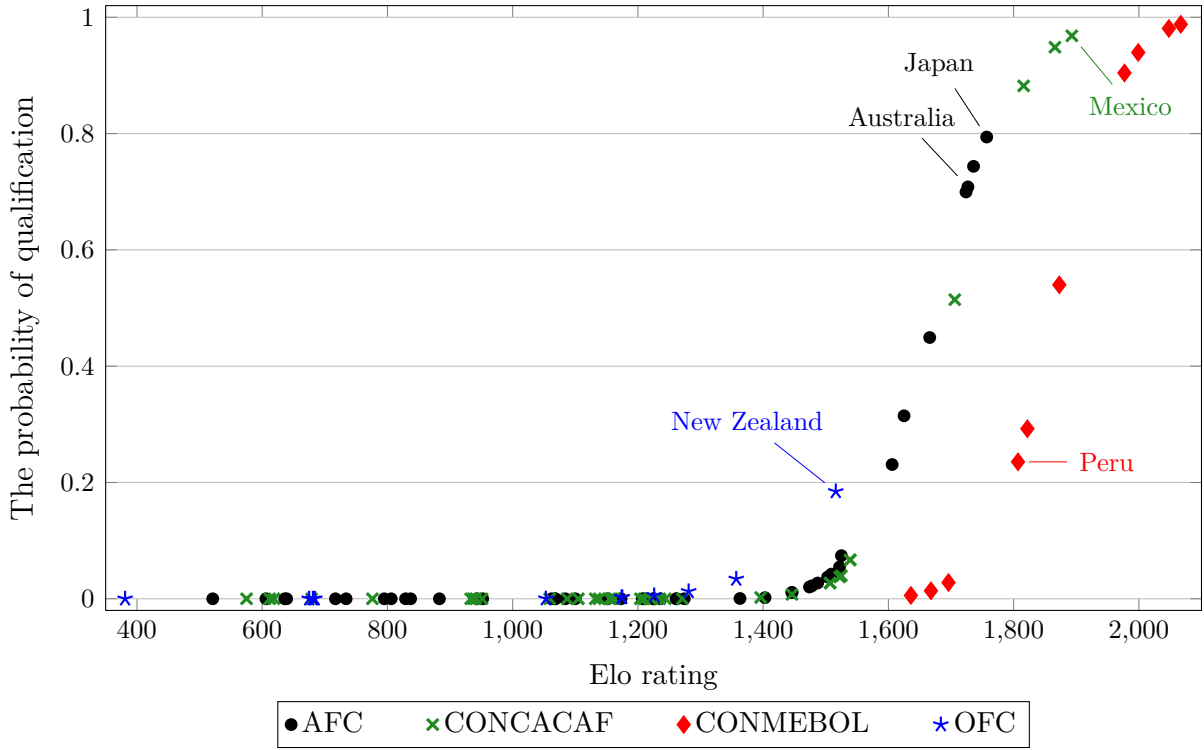


Figure~\ref{Fig1} shows the probability of qualification for the 2018 FIFA World Cup as the function of Elo ratings.
Unsurprisingly, the simple round-robin format of the CONMEBOL qualification guarantees that this tournament metric depends monotonically on the strength of the teams.
The structure of the OFC qualification does not necessarily satisfy the fairness condition but it still holds because only the four weakest teams should play in the first round and the seeding is based on the strengths of the teams. Similarly, the AFC and CONCACAF qualifications are essentially conforming to the principle of giving higher chances for better teams, too.

\begin{table}[t]
  \centering
  \caption{The level of unfairness within the confederations}
  \label{Table1}
  \rowcolors{1}{}{gray!20}
    \begin{tabularx}{0.5\textwidth}{lC} \toprule
    Confederation & Value of $\textit{UF}$ \\ \bottomrule
    AFC   & $6.23 \times 10^{-7}$ \\
    CONCACAF & $4.18 \times 10^{-6}$ \\
    CONMEBOL & 0 \\
    OFC   & 0 \\ \bottomrule
    \end{tabularx}
\end{table}

As Table~\ref{Table1} reveals, the qualification tournaments of all confederations are constructed fairly (the slot in the play-offs is counted as half). The negligible numbers for AFC and CONCACAF are only due to the stochastic nature of the simulation, which leads to volatile qualifying probabilities for weak teams. A team with a lower Elo rating never has a higher chance to qualify than a team with a higher Elo rating if either of these probabilities is above 0.002\%. On the other hand, some teams have the same Elo rating (see Tables~\ref{Table_A1} and \ref{Table_A2} in the Appendix), and these cases naturally increase $\textit{UF}$ according to Definition~\ref{Def1}.

Unfairness has another dimension, that is, between the confederations. According to Figure~\ref{Fig1}, the South American qualifiers seem to be the toughest. In order to investigate this issue, Peru (the $7$th strongest team in CONMEBOL, Elo: 1807) has been exchanged sequentially with the strongest teams in the other three confederations: Japan (AFC, Elo: 1757), Mexico (CONCACAF, Elo: 1893), and New Zealand (OFC, Elo: 1516). These countries are highlighted in Figure~\ref{Fig1}.

\begin{table}[t]
  \centering
  \caption{Qualifying probabilities when Peru is moved to another confederation}
  \label{Table2}
  \rowcolors{1}{gray!20}{}
    \begin{tabularx}{\textwidth}{ll CccC} \toprule \hiderowcolors
    Team  & Original & \multicolumn{4}{c}{Peru plays in} \\
          & confederation & AFC   & CONCACAF & CONMEBOL & OFC \\ \bottomrule \showrowcolors
    Japan & AFC   & 0.117 & 0.793 & 0.794 & 0.785 \\
    Mexico & CONCACAF & 0.968 & 0.508 & 0.968 & 0.966 \\
    Peru  & CONMEBOL & 0.887 & 0.877 & 0.235 & 0.455 \\
    New Zealand & OFC   & 0.185 & 0.183 & 0.185 & $5.67 \times 10^{-5}$ \\ \bottomrule
    \end{tabularx}
\end{table}

Table~\ref{Table2} reports the probabilities of qualification for the four nations if Peru would contest in various confederations. The numbers in the diagonal show that any team is the worst off when playing in the CONMEBOL qualifiers but the chance of Peru to participate in the 2018 FIFA World Cup would more than triple by playing in the AFC or CONCACAF zone. Being a member of the OFC would be less beneficial for Peru due to the lack of a direct qualification slot. Its effect can be seen to some extent in the qualifying probabilities of Japan and Mexico: since Peru would qualify with more than 97\% probability from the OFC qualifiers to the inter-confederation play-offs, the two teams would have a larger probability to face Peru there, which would reduce their chances to advance to the World Cup finals.

\subsection{A potential improvement of fairness}

The straightforward solution to handle unfairness between the confederations would be to reallocate the slots available for them, especially because the current allocation system lacks any statistical validation, does not ensure the qualification of the best teams in the world, and does not reflect the number of teams per federation \citep{StoneRod2016}. The whole process is far from being transparent and is mainly determined by political, cultural, and historical factors. Consequently, operations research has a limited role to influence the allocation of World Cup slots between the FIFA confederations.

However, the matching in the two inter-confederation play-offs is probably a variable to be chosen freely by the FIFA executives who are responsible for the tournament design, as illustrated by the two policies used recently. We have considered three possibilities:
\begin{itemize}
\item
\emph{Random draw} for the play-offs: the four participants from the confederations AFC, CONCACAF, CONMEBOL, and OFC are drawn randomly into two pairs;
\item
\emph{Fair draw} for the play-offs: the four participants are paired such that AFC vs.\ CONCACAF and CONMEBOL vs.\ OFC;
\item
\emph{Close draw} for the play-offs: the four participants are paired such that AFC vs.\ OFC and CONCACAF vs.\ CONMEBOL.
\end{itemize}

The random draw has been used in the \href{https://en.wikipedia.org/wiki/2006_FIFA_World_Cup_qualification}{2006 FIFA World Cup qualification}, as well as since the \href{https://en.wikipedia.org/wiki/2014_FIFA_World_Cup_qualification}{2014 FIFA World Cup qualifiers}. The close draw has been used in the \href{https://en.wikipedia.org/wiki/2010_FIFA_World_Cup_qualification}{2010 FIFA World Cup qualification competition}: it matches nations from closer time zones, which allows for better kick-off times. Therefore, it can be optimal for the players, as well as may maximise gate revenue and the value of television rights. Finally, the fair draw is inspired by Figure~\ref{Fig1} and Table~\ref{Table2} since the CONMEBOL team is usually the strongest and the OFC team is usually the weakest in the play-offs.

\begin{table}[t]
  \centering
  \caption{Unfairness and the draw for the play-offs}
  \label{Table3}
  
\begin{subtable}{\textwidth}
  \centering
  \caption{The overall level of unfairness}
  \label{Table3a}
    \begin{tabularx}{0.5\textwidth}{CCC} \toprule
    	\multicolumn{3}{c}{Draw policy for the play-offs} \\
    Random & Fair  & Close \\ \midrule
    0.01063 & 0.00985 & 0.01308 \\ \bottomrule
    \end{tabularx}
\end{subtable}

\vspace{0.5cm}
\begin{subtable}{\textwidth}
  \centering
  \caption{The qualifying probabilities of certain teams}
  \label{Table3b}
  \rowcolors{1}{gray!20}{}
    \begin{tabularx}{\textwidth}{llc CCC} \toprule \hiderowcolors   
    \multirow{2}[0]{*}{Team} & \multirow{2}[0]{*}{Confederation} & \multirow{2}[0]{*}{Elo rating} & \multicolumn{3}{c}{Draw policy for the play-offs} \\
          &       &       & Random & Fair  & Close \\ \bottomrule \showrowcolors
    Australia & AFC   & 1727  & 0.708 & 0.744 & 0.781 \\
    Japan & AFC   & 1757  & 0.744 & 0.778 & 0.810 \\
    Mexico & CONCACAF & 1893  & 0.968 & 0.975 & 0.958 \\
    Peru  & CONMEBOL & 1807  & 0.235 & 0.241 & 0.205 \\
    New Zealand & OFC   & 1516  & 0.184 & 0.041 & 0.179 \\ \toprule
    \end{tabularx}
\end{subtable}
\end{table}

The draws for the play-offs are compared in Table~\ref{Table3}, the measure of unfairness (Definition~\ref{Def1}) is presented in Table~\ref{Table3a}, while Table~\ref{Table3b} provides the probability of qualification for some countries. Again, there are some countries with the same Elo rating: Guatemala, Honduras (1539); Qatar, Trinidad and Tobago (1525); Jamaica, Oman (1522); Canada, El Salvador (1507); Haiti, Kuwait, North Korea (1446); Malaysia; Philippines (1225); India, Maldives (1068). These cases naturally increase the value of $\textit{UF}$, mostly when the national teams play in different confederations.

Intuitively, the fair draw is the least unfair. The close draw mostly favours AFC and OFC, however, it is detrimental to the members of the CONCACAF and CONMEBOL, implying the most severe unfairness.

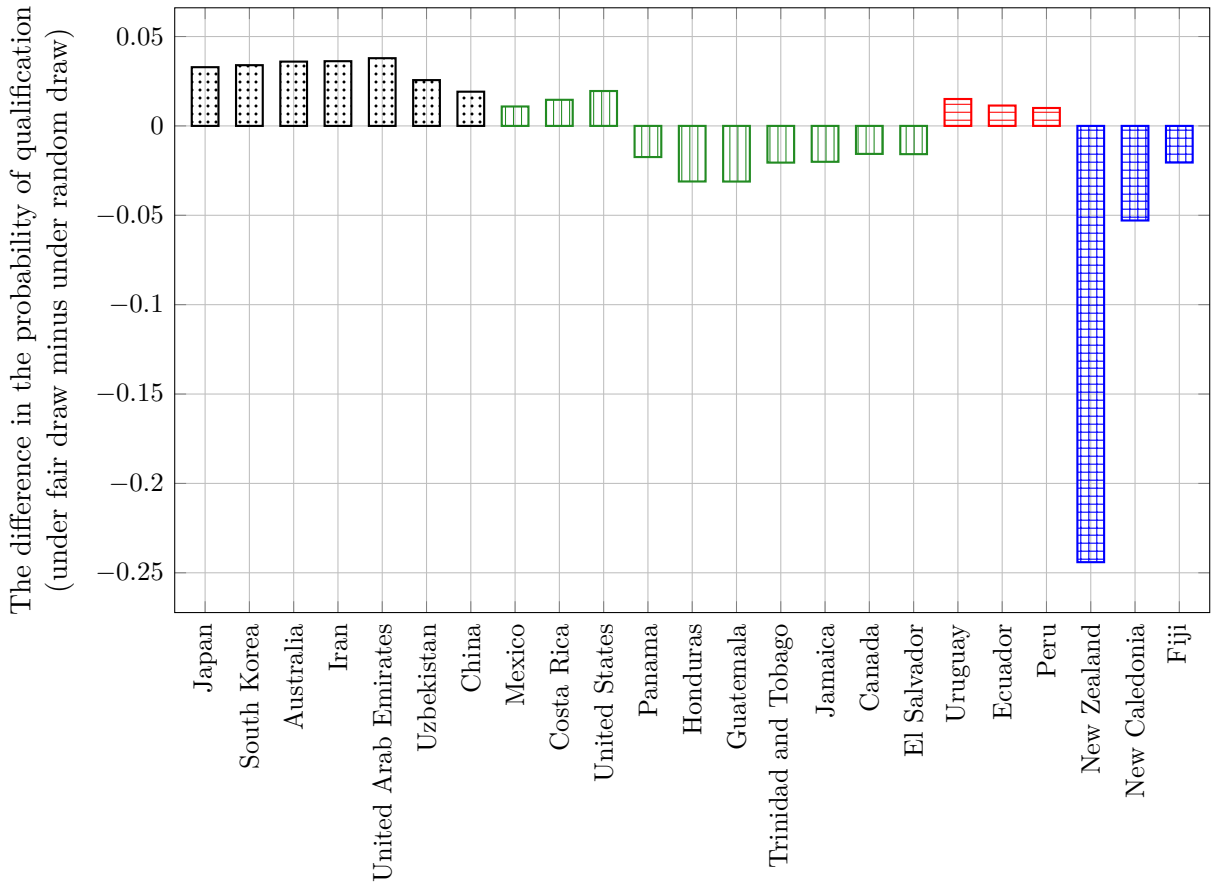
\begin{figure}[t]
\centering

\begin{tikzpicture}
\begin{axis}[width = 0.95\textwidth, 
height = 0.6\textwidth,
xmajorgrids,
ymajorgrids,
xlabel style = {font = \small},
symbolic x coords = {Japan, South Korea, Australia, Iran, United Arab Emirates, Uzbekistan, China, Mexico, Costa Rica, United States, Panama, Honduras, Guatemala, Trinidad and Tobago, Jamaica, Canada, El Salvador, Uruguay, Ecuador, Peru, New Zealand, New Caledonia, Fiji},
xtick = data,
x tick label style={rotate=90,anchor=east},
enlarge x limits = {abs = 0.4cm},
ybar stacked,
bar width = 10pt,
y tick label style={/pgf/number format/.cd,fixed,precision=3},
ylabel = {The difference in the probability of qualification \\ (under fair draw minus under random draw)},
ylabel style = {font = \small, align = center},
]
\addplot[black,thick,pattern=dots] coordinates {
(Japan,0.0328496)
(South Korea,0.0339748)
(Australia,0.0359712)
(Iran,0.036202)
(United Arab Emirates,0.0378812)
(Uzbekistan,0.0256228)
(China,0.019144)
(Mexico,0)
(Costa Rica,0)
(United States,0)
(Panama,0)
(Honduras,0)
(Guatemala,0)
(Trinidad and Tobago,0)
(Jamaica,0)
(Canada,0)
(El Salvador,0)
(Uruguay,0)
(Ecuador,0)
(Peru,0)
(New Zealand,0)
(New Caledonia,0)
(Fiji,0)
};

\addplot[ForestGreen,thick,pattern=vertical lines,pattern color=ForestGreen] coordinates {
(Japan,0)
(South Korea,0)
(Australia,0)
(Iran,0)
(United Arab Emirates,0)
(Uzbekistan,0)
(China,0)
(Mexico,0.0108623375668772)
(Costa Rica,0.0146015407368286)
(United States,0.0195181777407835)
(Panama,-0.0174246595480696)
(Honduras,-0.0310883368032467)
(Guatemala,-0.0311244084756179)
(Trinidad and Tobago,-0.0205336293479126)
(Jamaica,-0.0200776289613332)
(Canada,-0.0156619117850237)
(El Salvador,-0.0157946283154461)
(Uruguay,0)
(Ecuador,0)
(Peru,0)
(New Zealand,0)
(New Caledonia,0)
(Fiji,0)
};

\addplot[red,thick,pattern=horizontal lines,pattern color=red] coordinates {
(Japan,0)
(South Korea,0)
(Australia,0)
(Iran,0)
(United Arab Emirates,0)
(Uzbekistan,0)
(China,0)
(Mexico,0)
(Costa Rica,0)
(United States,0)
(Panama,0)
(Honduras,0)
(Guatemala,0)
(Trinidad and Tobago,0)
(Jamaica,0)
(Canada,0)
(El Salvador,0)
(Uruguay,0.0150507351474889)
(Ecuador,0.011391161895791)
(Peru,0.0100224801384608)
(New Zealand,0)
(New Caledonia,0)
(Fiji,0)
};

\addplot[blue,thick,pattern=grid,pattern color=blue] coordinates {
(Japan,0)
(South Korea,0)
(Australia,0)
(Iran,0)
(United Arab Emirates,0)
(Uzbekistan,0)
(China,0)
(Mexico,0)
(Costa Rica,0)
(United States,0)
(Panama,0)
(Honduras,0)
(Guatemala,0)
(Trinidad and Tobago,0)
(Jamaica,0)
(Canada,0)
(El Salvador,0)
(Uruguay,0)
(Ecuador,0)
(Peru,0)
(New Zealand,-0.244054809885272)
(New Caledonia,-0.0529089761974604)
(Fiji,-0.020485987516479)
};
\end{axis}
\end{tikzpicture}

\captionsetup{justification=centering}
\caption[The effect of the reform]{The effect of improving the fairness of the 2018 FIFA World Cup qualification \\ via modifying the draw for inter-confederation play-offs}
\label{Fig2}
\end{figure}


The effect of a fair draw is detailed in Figure~\ref{Fig2} for the teams with at least one percentage point change in the probability of qualification to the 2018 FIFA World Cup. Compared to the current random design, all Asian and South American countries would be better off and the strongest CONCACAF countries are also preferred. On the other hand, all nations of the OFC, in particular, the dominating New Zealand, would lose substantially from this reform. The gains are distributed more equally because there is no such a prominent team in the other zones. Some weaker CONCACAF members are worse off due to the impossibility of playing against New Zealand in the inter-confederation play-offs.

\subsection{Counterfactual: was it favourable for Australia to join the AFC?}

FIFA president \emph{Sepp Blatter} had promised a full slot to the OFC as part of his re-election campaign in November 2002 but the suggestion was reconsidered in June 2003 \citep{ABC2003}.
Subsequently, the largest and most successful nation of the OFC, Australia, left to join the AFC in 2006. It raises the interesting question of how this move has affected the 2018 FIFA World Cup qualification.

First, the unfairness measure $\textit{UF}$ would be $0.013$ with Australia playing in the OFC, which corresponds to an increase of more than 20\% compared to the current situation. The action of Australia has contributed to the fairness of the 2018 FIFA World Cup qualification. Furthermore, the magnitude of the improvement is higher than the gain from the proposed novel draw for the inter-confederation play-offs. This reduction in unfairness likely explains why the move of Australia was approved.

\begin{figure}[t]
\centering

\begin{tikzpicture}
\begin{axis}[width = 0.96\textwidth, 
height = 0.6\textwidth,
align = center,
xmajorgrids,
ymajorgrids,
xlabel style = {font = \small},
symbolic x coords = {Japan, South Korea, Australia, Iran, United Arab Emirates, Uzbekistan, China, Qatar, Oman, Iraq, Jordan, Bahrain, Syria, Saudi Arabia, Panama, Honduras, Guatemala, Trinidad and Tobago, Jamaica, Canada, El Salvador, Uruguay, Ecuador, Peru, New Zealand, New Caledonia, Fiji},
xtick = data,
x tick label style={rotate=90,anchor=east},
enlarge x limits = {abs = 0.4cm},
ybar stacked,
bar width = 8pt,
y tick label style={/pgf/number format/.cd,fixed,precision=3},
ylabel = {The difference in the probability of qualification \\ (Australia in the AFC minus Australia in the OFC)},
ylabel style = {font = \small, align = center},
]
\addplot[black,thick,pattern=dots] coordinates {
(Japan,-0.0746348)
(South Korea,-0.08437)
(Australia,0.2774316)
(Iran,-0.0811676)
(United Arab Emirates,-0.1642724)
(Uzbekistan,-0.0864628)
(China,-0.1015284)
(Qatar,-0.0333468)
(Oman,-0.0457604)
(Iraq,-0.0241792)
(Jordan,-0.0224912)
(Bahrain,-0.0169664)
(Syria,-0.0144888)
(Saudi Arabia,-0.0133328)
(Panama,0)
(Honduras,0)
(Guatemala,0)
(Trinidad and Tobago,0)
(Jamaica,0)
(Canada,0)
(El Salvador,0)
(Uruguay,0)
(Ecuador,0)
(Peru,0)
(New Zealand,0)
(New Caledonia,0)
(Fiji,0)
};

\addplot[ForestGreen,thick,pattern=vertical lines,pattern color=ForestGreen] coordinates {
(Japan,0)
(South Korea,0)
(Australia,0)
(Iran,0)
(United Arab Emirates,0)
(Uzbekistan,0)
(China,0)
(Qatar,0)
(Oman,0)
(Iraq,0)
(Jordan,0)
(Bahrain,0)
(Syria,0)
(Saudi Arabia,0)
(Panama,0.0563908)
(Honduras,0.0242056)
(Guatemala,0.0243888)
(Trinidad and Tobago,0.0157644)
(Jamaica,0.015392)
(Canada,0.011768)
(El Salvador,0.0117632)
(Uruguay,0)
(Ecuador,0)
(Peru,0)
(New Zealand,0)
(New Caledonia,0)
(Fiji,0)
};

\addplot[red,thick,pattern=horizontal lines,pattern color=red] coordinates {
(Japan,0)
(South Korea,0)
(Australia,0)
(Iran,0)
(United Arab Emirates,0)
(Uzbekistan,0)
(China,0)
(Qatar,0)
(Oman,0)
(Iraq,0)
(Jordan,0)
(Bahrain,0)
(Syria,0)
(Saudi Arabia,0)
(Panama,0)
(Honduras,0)
(Guatemala,0)
(Trinidad and Tobago,0)
(Jamaica,0)
(Canada,0)
(El Salvador,0)
(Uruguay,0.0330632)
(Ecuador,0.02983)
(Peru,0.0257156)
(New Zealand,0)
(New Caledonia,0)
(Fiji,0)
};

\addplot[blue,thick,pattern=grid,pattern color=blue] coordinates {
(Japan,0)
(South Korea,0)
(Australia,0)
(Iran,0)
(United Arab Emirates,0)
(Uzbekistan,0)
(China,0)
(Qatar,0)
(Oman,0)
(Iraq,0)
(Jordan,0)
(Bahrain,0)
(Syria,0)
(Saudi Arabia,0)
(Panama,0)
(Honduras,0)
(Guatemala,0)
(Trinidad and Tobago,0)
(Jamaica,0)
(Canada,0)
(El Salvador,0)
(Uruguay,0)
(Ecuador,0)
(Peru,0)
(New Zealand,0.1586464)
(New Caledonia,0.0332156)
(Fiji,0.0125892)
};
\end{axis}
\end{tikzpicture}

\captionsetup{justification=centering}
\caption[The effect of the reform]{The effect of Australia being a member of the AFC \\ on the 2018 FIFA World Cup qualification}
\label{Fig3}

\end{figure}
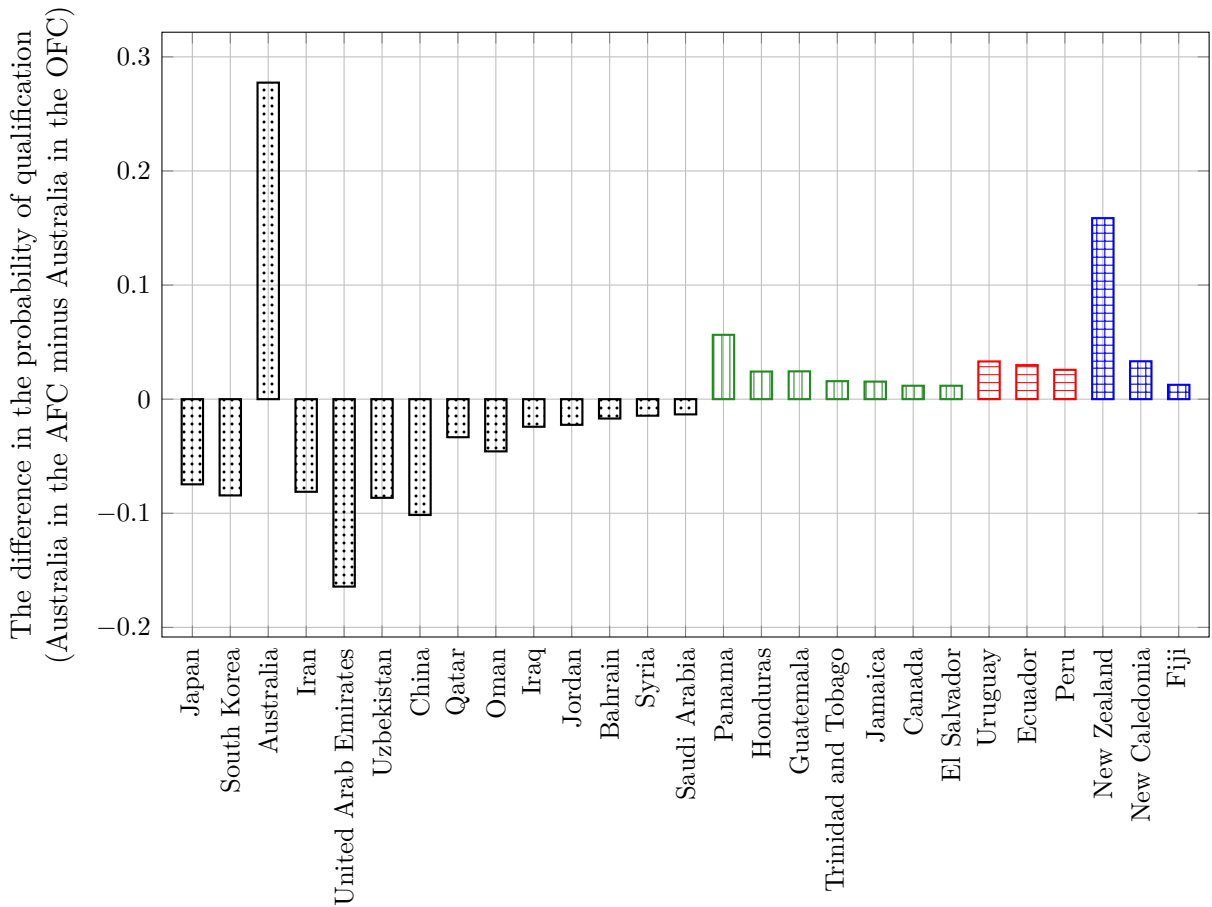


Second, the probabilities of qualification are computed if Australia would have remained in the OFC. Figure~\ref{Fig3} plots the effects for the national teams facing a change of at least one percentage point. Notably, Australia has increased the probability of participating in the 2018 FIFA World Cup from 43\% to 70\% by leaving the OFC for the AFC. The move has also been strongly favourable for New Zealand, which is now the strongest OFC team and has more than 70\% chance to grab the slot guaranteed in the play-offs for Oceania. Every CONCACAF and CONMEBOL member has been better off due to the reduction in the expected strength of the countries contesting in the play-offs. However, all original AFC nations have lost with the entrance of Australia, especially those teams that are only marginally weaker than Australia.

\section{Conclusions} \label{Sec6}

We have analysed four series of qualification tournaments for the 2018 FIFA World Cup via Monte Carlo simulations. Their design does not suffer from serious problems.
However, there are substantial differences between the chances of nations playing in different continents: Peru would have tripled its probability of qualification by competing in the AFC or CONCACAF zone, whereas New Zealand would have lost any prospect of participation by being a member of the CONMEBOL. Australia is found to have greatly benefited from leaving the OFC for the AFC in 2006.

In addition, probably first in the literature, a measure of unfairness has been proposed to quantify to which extent weaker teams are preferred over stronger teams by a tournament design. A simple modification in the format of the inter-confederation play-offs can reduce this metric by about 7.5\%, hence FIFA has a good reason to consider the novel policy.

Hopefully, this paper will become only a pioneering attempt of academic researchers to study the qualification to the soccer World Cup. There remains huge scope for improving the simulation model, especially concerning the prediction of individual matches. Nonetheless, even these results might be useful for sports governing bodies: we believe that FIFA could further increase the economic success of World Cups by using a more transparent and statistically validated method in the allocation of qualifying berths and the design of confederation-level qualification tournaments.

\section*{Acknowledgements}
\addcontentsline{toc}{section}{Acknowledgements}
\noindent
This paper could not have been written without \emph{my father} (also called \emph{L\'aszl\'o Csat\'o}), who has primarily coded the simulations in Python. \\
We are grateful to \emph{Mark Gagolewski}, \emph{Dries Goossens}, \emph{Lars Hvattum}, and \emph{Cristophe Ley} for useful advice. \\
Five anonymous reviewers provided valuable comments and suggestions on earlier drafts. \\
We are indebted to the \href{https://en.wikipedia.org/wiki/Wikipedia_community}{Wikipedia community} for collecting and structuring valuable information on the sports tournaments discussed. \\
The research was supported by the MTA Premium Postdoctoral Research Program grant PPD2019-9/2019.

\bibliographystyle{apalike}
\bibliography{All_references}

\clearpage
\section*{Appendix}
\addcontentsline{toc}{section}{Appendix}

\renewcommand\thetable{A.\arabic{table}}
\setcounter{table}{0}

\makeatletter
\renewcommand\p@subtable{A.\arabic{table}}
\makeatother

\renewcommand\thefigure{A.\arabic{figure}}
\setcounter{figure}{0}

\makeatletter
\renewcommand\p@subfigure{A.\arabic{figure}}
\makeatother

\begin{table}[ht!]
  \centering
  \caption{CONCACAF nations participating in the 2018 FIFA World Cup qualification}
  \label{Table_A1}
\begin{threeparttable}
  \rowcolors{1}{}{gray!20}  
    \begin{tabularx}{0.75\textwidth}{lC} \toprule \hiderowcolors
    Country & Elo rating of 21 March 2015 \\ \bottomrule \showrowcolors   
    \textbf{Costa Rica} & 1866 \\
    \textbf{Mexico} & 1893 \\
    United States & 1816 \\
    \emph{Honduras} & 1539 \\
    \textbf{Panama} & 1706 \\
    Trinidad and Tobago & 1525 \\
    Jamaica & 1522 \\
    Haiti & 1446 \\ \hline
    Canada & 1507 \\
    Cuba  & 1396 \\
    Aruba & 941 \\
    Dominican Republic & 1235 \\
    El Salvador & 1507 \\
    Suriname & 1244 \\
    Guatemala & 1539 \\
    Saint Vincent and the Grenadines & 1162 \\
    Saint Lucia & 1132 \\
    Grenada & 1158 \\
    Antigua and Barbuda & 1271 \\
    Guyana & 1221 \\
    Puerto Rico & 1059 \\ \hline
    Saint Kitts and Nevis & 1204 \\
    Belize & 1138 \\
    Montserrat & 619 \\
    Dominica & 951 \\
    Barbados & 1155 \\
    Bermuda & 1210 \\
    Nicaragua & 1105 \\
    Turks and Caicos Islands & 776 \\
    Curacao & 1087 \\
    U.S. Virgin Islands & 679 \\
    Bahamas & 933 \\
    Cayman Islands & 937 \\
    British Virgin Islands & 612 \\
    Anguilla & 575 \\ \toprule
    \end{tabularx}
\begin{tablenotes} \footnotesize
\item
The countries are ranked according to the FIFA World Ranking of August 2014.
The first match in the first round was played on 22 March 2015.
\item
The team(s) written in \textbf{bold} (\emph{italics}) qualified for the 2018 FIFA World Cup (inter-confederation play-offs).
\end{tablenotes}
\end{threeparttable}
\end{table}

\begin{table}[ht!]
  \centering
  \caption{AFC nations participating in the 2018 FIFA World Cup qualification}
  \renewcommand\arraystretch{0.88}
  \label{Table_A2}
\begin{threeparttable}
  \rowcolors{1}{gray!20}{}  
    \begin{tabularx}{0.65\textwidth}{lC} \toprule \hiderowcolors
    Country & Elo rating of 11 March 2015 \\ \bottomrule \showrowcolors  
    \textbf{Iran}  & 1724 \\
    \textbf{Japan} & 1757 \\
    \textbf{South Korea} & 1736 \\
    \textbf{\emph{Australia}} & 1727 \\
    United Arab Emirates & 1666 \\
    Uzbekistan & 1625 \\
    China & 1606 \\
    Iraq  & 1509 \\
    \textbf{Saudi Arabia} & 1474 \\
    Oman  & 1522 \\
    Qatar & 1525 \\
    Jordan & 1503 \\
    Bahrain & 1487 \\
    Vietnam & 1236 \\
    Syria & 1478 \\
    Kuwait & 1446 \\
    Afghanistan & 1084 \\
    Philippines & 1225 \\
    Palestine & 1274 \\
    Maldives & 1068 \\
    Thailand & 1363 \\
    Tajikistan & 1262 \\
    Lebanon & 1403 \\
    Kyrgyzstan & 1100 \\
    North Korea & 1446 \\
    Myanmar & 1062 \\
    Turkmenistan & 1261 \\
    Indonesia & 1207 \\
    Singapore & 1173 \\
    Malaysia & 1225 \\
    Hong Kong & 1152 \\
    Bangladesh & 952 \\
    Guam  & 806 \\
    Laos  & 837 \\ \hline
    India & 1068 \\
    Sri Lanka & 829 \\
    Yemen & 1211 \\
    Cambodia & 717 \\
    Chinese Taipei & 795 \\
    Timor-Leste & 639 \\
    Nepal & 883 \\
    Macau & 636 \\
    Pakistan & 943 \\
    Mongolia & 734 \\
    Brunei & 606 \\
    Bhutan & 521 \\ \toprule
    \end{tabularx}
\begin{tablenotes} \footnotesize
\item
The countries are ranked according to the FIFA World Ranking of January 2015 ($35$--$46$) and April 2015 ($1$--$34$).
The first match in the first round was played on 12 March 2015.
\item
The team(s) written in \textbf{bold} (\emph{italics}) qualified for the 2018 FIFA World Cup (inter-confederation play-offs).
\end{tablenotes}
\end{threeparttable}
\end{table}

\begin{table}[ht!]
  \centering
  \caption{CONMEBOL nations participating in the 2018 FIFA World Cup qualification}
  \label{Table_A3}
\begin{threeparttable}
  \rowcolors{1}{gray!20}{}  
    \begin{tabularx}{0.6\textwidth}{lC} \toprule \hiderowcolors
    Country & Elo rating of 7 October 2015 \\ \bottomrule \showrowcolors    
    \textbf{Argentina} & 2067 \\
    \textbf{Colombia} & 1999 \\
    \textbf{Brazil} & 2048 \\
    Chile & 1977 \\
    \textbf{Uruguay} & 1873 \\
    Ecuador & 1822 \\
    \textbf{\emph{Peru}}  & 1807 \\
    Paraguay & 1696 \\
    Bolivia & 1636 \\
    Venezuela & 1668 \\ \toprule
    \end{tabularx}
\begin{tablenotes} \footnotesize
\item
The countries are ranked according to the FIFA World Ranking of October 2015. The first matchday was 8 October 2015.
\item
The team(s) written in \textbf{bold} (\emph{italics}) qualified for the 2018 FIFA World Cup (inter-confederation play-offs).
\end{tablenotes}
\end{threeparttable}
\end{table}

\begin{table}[ht!]
  \centering
  \caption{OFC nations participating in the 2018 FIFA World Cup qualification}
  \label{Table_A4}
\begin{threeparttable}
  \rowcolors{1}{gray!20}{}  
    \begin{tabularx}{0.6\textwidth}{lC} \toprule \hiderowcolors
    Country & Elo rating of 30 August 2015 \\ \bottomrule \showrowcolors   
    \emph{New Zealand} & 1516 \\
    New Caledonia & 1357 \\
    Tahiti & 1226 \\
    Solomon Islands & 1175 \\
    Vanuatu & 1174 \\
    Fiji  & 1281 \\
    Papua New Guinea & 1053 \\ \hline
    Samoa & 681 \\
    Tonga & 684 \\
    American Samoa & 381 \\
    Cook Islands & 675 \\ \bottomrule
    \end{tabularx}
\begin{tablenotes} \footnotesize
\item
The countries are ranked according to the FIFA World Ranking of July 2015, separately for the positions $1$--$7$ (byes in the first round) and $8$--$11$ (playing in the first round).
The first match in the first round was played on 31 August 2015.
\item
The team written in \emph{italics} qualified for the inter-confederation play-offs of the 2018 FIFA World Cup.
\end{tablenotes}
\end{threeparttable}
\end{table}

\end{document}